\documentclass[aps,pra,twocolumn,superscriptaddress,
nofootinbib]{revtex4-1}
\usepackage{mathrsfs}
\usepackage{amsfonts}
\usepackage{amstext}
\usepackage{amsmath}
\usepackage{amssymb}
\usepackage{bm}
\usepackage{dcolumn}
\usepackage{bbm}
\usepackage[dvips]{graphicx}
\def\qed{\leavevmode\unskip\penalty9999 \hbox{}\nobreak\hfill
     \quad\hbox{\leavevmode  \hbox to.77778em{%
              \hfil\vrule   \vbox to.675em%
               {\hrule width.6em\vfil\hrule}\vrule\hfil}}
     \par\vskip3pt}

\def\ra{\rangle}
\def\la{\langle}

\begin{document}

\title{Discordlike correlation of bipartite coherence}

\author{Yu Guo}
\email{guoyu3@aliyun.com}
\affiliation{Institute of Quantum Information Science, Shanxi Datong University, Datong, Shanxi 037009, China}
\affiliation{Institute for Quantum Science and Technology, 
	University of Calgary, Alberta T2N 1N4, Canada}

\author{Sumit Goswami}
\affiliation{Institute for Quantum Science and Technology, 
	University of Calgary, Alberta T2N 1N4, Canada}%

\begin{abstract}

Quantum discord has been studied extensively as a measure of non-classical
correlations which includes entanglement as a subset. Although it is well
known that non-zero discord can exist without entanglement, the origin of
quantum discord is not well understood as compared to entanglement, which
manifests itself more simply as inseparable higher dimensional quantum superposition.
In this paper we establish the discordlike correlation of bipartite
coherence and then compare it to quantum discord. Consequently, we show
that the minimum of the discordlike correlation of coherence coincides
with the original quantum discord. This demonstrates quantum discord as
the irreducible correlated bipartite coherence. In addition, the discordlike
correlated coherence is shown to admit the postulates of the quantum
resource theory (QRT), although the original quantum discord is not
a ``good'' candidate under the QRT. We also find that the relative
entropy measure induced from the discordlike coherence is a
well-defined coherence measure for bipartite states.

\end{abstract}

\maketitle

\section{Introduction}

Correlated information always lies at the heart of quantum information
theory \cite{Nielsen,Horodecki2009,Guhne}.
Quantum discord was introduced in \cite{Ollivier} to quantify the total amount
of quantum correlation present in a bipartite system. Entanglement,
the most widely used quantum correlation, is included as a subset in quantum discord.
It is shown that quantum discord is more robust against
environment induced decoherence than entanglement \cite{Werlang,Xujinshi}.
Moreover, it has been proven to be an important
quantum resource in a plethora of quantum information processing tasks
~\cite{Modi2012,Streltsov2013,Yang,Streltsov2011prl,Huxueyuan2012pra,
Guo2016srep,guo2013jpa-qd,Datta,Brodutch2013,Suxiaolong2014,Dakic12,
Giorgi2013pra,Pirandola,Xu2013nat,Franco2012pra,Humingliang2014aop,Humingliang2015aop}.

The relation between different quantum resources is of great
importance \cite{Luo2016pra,Streltsov2015prl,Tan}. It is shown
that entanglement is a minimal quantum discord over state
extensions \cite{Luo2016pra}. Another fundamental concept in quantum
physics which is closely connected to quantum superposition and
quantum correlations is coherence. An algorithmic characterization of
quantum coherence as a resource and a set of bona fide criteria for
coherence monotones have been identified \cite{Baumgratz2014prl,Girolami,
Dushuanping2015ap,Chenjianxin2016pra,Napoli2016prl,Rastegin2016pra,
Rana2016pra,Streltsov2015prl,Shaolianhe2015pra,Dushuanping2015pra,Humingliang2016srep}.
Correlated coherence has been proposed to capture the mutual coherence
between the two subsystems of a bipartite system \cite{Tan}.
It is shown that coherence can be measured with entanglement
\cite{Streltsov2015prl}, the correlated coherence is closely related
to entanglement \cite{Tan} and the basis-free relative entropy of
coherence coincides with the relative entropy of the quantum
discord \cite{Yaoyao}. The main purpose of this paper is to investigate
a discordlike correlation of bipartite coherence and then compare it
to the original quantum discord. By replacing the von Neumann entropy
with the relative entropy measure of coherence, replacing the mutual
information by the correlated coherence and replacing the von Neumann
measurements by the local rank-one projective physically incoherent operations
we establish a discordlike correlation of coherence for bipartite states
and then compare it to the original quantum discord: the original quantum
discord turned out to be the minimal discordlike correlation of coherence
over all possible reference bases.

Later we investigate the role of the discordlike correlation of
coherence as quantum resources in context of the recently developed
quantum resource theory (QRT) \cite{Gilad2015prl}. QRT was developed
to create a unifying theoretical framework for different quantum resources.
Considerable work on formulating QRT has been done recently
\cite{Brandao2013prl,Horodecki2013nat,Faist2015nat,Gour2015pr,Marvian2014nat,
Braunstein2005rmp,Grudka2014prl,Rivas2014rpp,Gilad2015prl,
Winter2016prl,Liuziwen2017prl}.
A general structure of QRT has three ingredients: (1) the free states,
(2) the resource states, and (3) the free operations.
For example, in entanglement theory the resource states are the entangled states,
the free states are the separable ones and the free operations
are the local operations and classical communication
(LOCC). In the theory of coherence, the resource states are the coherent states,
the free states are the incoherent ones and the free operations
are the incoherent operations. However, not all the free operations can be
implemented physically. Hence, the physically consistent
conditions for QRT was formulated very recently \cite{Gilad2016prl},
particularly in the context of coherence. Physically incoherent operations
turned out to be some special incoherent operations \cite{Gilad2016prl}.
Consequently, we show that although the original quantum discord does not
obey the structure of QRT, the discordlike correlation of
coherence demonstrated itself as a reasonable resource under QRT.

The rest of this paper is structured as follows. In Sec. II,
we give a brief overview of the quantum discord and the
coherence. The discordlike correlation of coherence is established in Sec. III
and then we obtain a relative entropy of the discordlike correlation in Sec. IV.
Section V discuss the discordlike correlation as a quantum resource under QRT.
We conclude in Sec. VI.

\section{Preliminary notions}

We recall the definitions of quantum discord and coherence at first.
For a state $\rho_{ab}$ of a bipartite system, with finite dimensional
subsystems $A$ and $B$, described by Hilbert space $H_a\otimes H_b$, the
quantum discord of $\rho_{ab}$ (up to part $A$) is defined by \cite{Ollivier}
\begin{eqnarray}
D^a(\rho_{ab}):=\min_{\Pi^a}\{I(\rho_{ab})-I(\rho_{ab}|\Pi^a)\},\label{qd}
\end{eqnarray}
where, the minimum is taken over all local von Neumann measurements
$\Pi^a$ [i.e., $\Pi^a(\cdot)=\sum_i\Pi_i^a\otimes I_b(\cdot)\Pi_i^a\otimes I_b$
with $\Pi_i^a=|\psi_i\ra\la\psi_i|$ for some orthonormal basis $\{|\psi_i\ra\}$ of $H_a$],
$I(\rho_{ab}):=S(\rho_a)+S(\rho_b)-S(\rho_{ab})$
is interpreted as the quantum mutual information, $S(\rho):=-{\rm Tr}(\rho\log_2\rho)$
is the von Neumann entropy, and $I(\rho_{ab}|\Pi^a):=S(\rho_b)-\sum_kp_kS(\rho_k)$ with
$p_k\rho_k=(\Pi_k^a\otimes I_b)\rho_{ab}(\Pi_k^a\otimes I_b)$,
$k=1$, 2, $\dots$, $N$, $1\leq N\leq\dim H_a$. $D^a(\rho_{ab})$ can be quantified as
\begin{eqnarray}
D^a(\rho_{ab})=\min_{\Pi^a}\{I(\rho_{ab})-I[(\Pi^a\otimes \mathbbm{1}_b)\rho_{ab}]\}.\label{qd2}
\end{eqnarray}
Here $I_a$ and $I_b$ are the identity operators on $H_a$ and $H_b$, respectively, and $\openone_{b}$ denotes the
identity map on part $B$.

Coherence is defined and quantified along the approach in
~\cite{Baumgratz2014prl}. Let $H$ be a finite-dimensional Hilbert space with
$\dim H=d$. Fixing a particular basis $\{|i\rangle\}_{i=0}^{d-1}$,
we call all quantum states represented by density operators that are
diagonal in this basis incoherent. This incoherent set
of quantum states will be labeled by $\mathcal{I}$, all density
operators $\rho\in \mathcal{I}$ are of the form
\begin{equation}
\rho=\sum_{i=0}^{d-1}\delta_i|i\rangle\langle i|.
\end{equation}
Henceforth, we call the fixed basis reference basis. If $\{|i\ra\}$ and
$\{|j'\ra\}$ are reference bases of $H_a$ and $H_b$ respectively,
then $\{|i\ra|j'\ra\}$ is the reference basis of $H_a\otimes H_b$.
A quantum operation is incoherent if its Kraus operators fulfill
$K_n\rho K_n^\dag/{\rm Tr}(K_n\rho K_n^\dag)\in \mathcal{I}$
for all $\rho\in \mathcal{I}$ and for all $n$.
For a state space $H$ we denote by $\mathcal{B}(H)$ and
$\mathcal{S}(H)$ the space of all bounded linear operators on
$H$ and the set of all quantum states on $H$, respectively.

\section{Discordlike correlation of bipartite coherence}

The relative entropy of coherence is defined by~\cite{Baumgratz2014prl}
\begin{eqnarray}
C_r(\rho):=\min\limits_{\sigma\in\mathcal{I}}S(\rho\|\sigma),
\end{eqnarray}
where $S(\rho\|\sigma)={\rm Tr}(\rho\log_2\rho-\rho\log_2\sigma)$ is the relative entropy.
$C_r(\rho)$ can be calculated to be ~\cite{Baumgratz2014prl}
\begin{eqnarray}
C_r(\rho)=S[\Delta(\rho)]-S(\rho),
\end{eqnarray}
where $\Delta(\rho)$ denotes the diagonal part of the $\rho$ in the reference basis,
i.e.,
\begin{eqnarray}
\Delta(\rho)=\sum_i|i\ra\la i|\rho|i\ra\la i|.
\end{eqnarray}
$C_r$ admits the supper-additive property \cite{Xi}
\begin{eqnarray}
C_r(\rho_{ab})\geq C_r(\rho_a)+C_r(\rho_b).
\label{supadditive}
\end{eqnarray}
The equality holds whenever $\rho_{ab}$ is a product state \cite{Xi}.
We remark here that $S(\rho_{ab})\leq S(\rho_a)+S(\rho_b)$ and the
equality holds if and only if $\rho_{ab}$ is a product state.
However, $C_r(\rho_{ab})=C_r(\rho_a)+C_r(\rho_b)$
whenever $\rho_{ab}$ is a product state but not vice versa.
In fact, for any diagonal bipartite state $\rho_{ab}$,
we have $C_r(\rho_{ab})=C_r(\rho_a)+C_r(\rho_b)$.
But $\rho_{ab}$ is not necessarily a product state.
If $\rho_{ab}$ is a maximally coherent state,
the equality in Eq.~(\ref{supadditive}) holds if and only
if it is a product state, provided that $\dim H_a=\dim H_b$.
But it is not valid when $\dim H_a\neq \dim H_b$ \cite{Baizhaofang2015qic}.

\begin{table*}
\caption{\label{tab:table1}The comparison of the discordlike correlation of coherence $D^a_c$
with the original quantum discord $D^a$.}
\begin{ruledtabular}
\begin{tabular}{cccccc}
 Correlation& Free states  & Physically free operation
&Invariant operation & Measurement \\ \hline
 ${D}^a$&$\mathcal{D}_0^a$&$\mathcal{E}_{u}\otimes\mathcal{E}_b$\footnotemark[1]&Local UO & $\Pi^a$\\
 ${D}_c^a$&$\mathcal{D}_{c-0}^a$&$\mathcal{E}_{iuo}\otimes\mathcal{E}_b$\footnotemark[2]
  &Local IUO& $\check{\Pi}^a$
 \\
\end{tabular}
\end{ruledtabular}
\footnotetext[1]{$\mathcal{E}_{u}$ denotes the unitary operation,
$\mathcal{E}_b$ is any given local operation on part $B$.}
\footnotetext[2]{$\mathcal{E}_{iuo}$ denotes the incoherently unitary operation.}
\end{table*}

{\it Definition 1.} For any $\rho_{ab}\in\mathcal{S}(H_a\otimes H_b)$,
we call the difference
\begin{eqnarray}
{I}_{co}(\rho_{ab}):={C}_r(\rho_{ab})
-{C}_r(\rho_a)-{C}_r(\rho_b)
\label{mutualcoherence}
\end{eqnarray}
the \emph{correlated coherence} of $\rho_{ab}$ with respect
to the relative entropy measure of coherence.

By definition, ${I}_{co}(\rho_{ab})$ is the amount of
mutual coherence information contained in $\rho_{ab}$,
which is similar to that of mutual information $I(\rho_{ab})$.
We remark here that the correlated coherence
$\mathcal{C}_{cc}(\rho_{ab}):=\mathcal{C}(\rho_{ab})
-\mathcal{C}(\rho_a)-\mathcal{C}(\rho_b)$ was proposed first
by the authors of~\cite{Tan}, where $\mathcal{C}$ is the $l_1$ norm measure of coherence,
i.e., $\mathcal{C}(\rho):=\sum_{i\neq j}|\la i|\rho|j\ra|$
with respect to the reference basis $\{|i\ra\}$.
[We use $C_r$ instead of the $l_1$ norm here since, as will be shown,
$I_{co}$ can induce the discordlike correlation (Definition 2) that
connects with the original quantum discord closely (Theorem 2).
However, one can easily check that the $l_1$ norm can not reveal such a relation.]

In~\cite{Gilad2016prl}, the \emph{physically
incoherent operation} (PIO) is proposed according to the
physically consistent QRT. A completely positive trace preserving (CPTP) map
is proven to be a PIO if and only if it can be written as
a convex combination of operations each with Kraus operators
$\{K_j\}_{j=1}^r$ of the form
\begin{eqnarray}
K_j=U_jP_j=\sum\limits_ye^{i\theta_y}|\pi_j(y)\rangle\langle y|P_j,
\label{pio}
\end{eqnarray}
where $\{|y\rangle\}$ is the reference basis of part $A$,
$P_j$ forms an orthogonal and complete set of incoherent projectors on part $A$,
$\sum_{j=1}^rK_j^\dag K_j=I$
and $\pi_j$ are permutations.
Namely, a quantum operation $\mathcal{E}$ is a PIO if and only if it
can be written as $\mathcal{E}(\cdot)=\sum_{i=1}^mt_i\mathcal{E}_i(\cdot)$
with $\sum_it_i=1$, $t_i>0$, $m\geq1$. Here
$\mathcal{E}_i(\cdot)=\sum_jK_j^{(i)}(\cdot){K_j^{(i)}}^\dag$
with $K_j^{(i)}$s as in Eq.~(\ref{pio}).

Particularly, we call a PIO as projective PIO (PPIO) if $m =1$, i.e.,
$\mathcal{E}$ is a PPIO if and only if $\mathcal{E}(\cdot)=\sum_jK_j(\cdot)K_j^\dag$
where $K_j$s admit the form in Eq.~(\ref{pio}). In other words, any PIO is a convex
combination of PPIOs. In a PPIO there is only one set of Kraus
operators and  the action of these Kraus operators $K_j$ are
similar to that of the projective operators $P_j$ (as $K_j =U_jP_j$
is unitarily related to $P_j$). Hence, we call this special
kind of PIO as ``projective'' PIO.
Here we note that $U_j$ and $P_j$ are defined with respect to a
fixed reference basis (that of coherence), as they are defined
as incoherent unitary and incoherent projector, respectively. For
the fixed reference basis $\{|y\rangle\}$ the form of $U_j$ is
given in Eq.~(\ref{pio}) and $\{P_j\}$ are any orthogonal and
complete set of projectors in that basis.

Moreover, if all the projectors ($P_j$) of a PPIO are rank-one,
we also have all Kraus operators (related by the incoherent unitary)
of rank one. We define this special case of a PPIO as a rank-one PPIO.
For example, in the case of a three-dimensional Hilbert space with
reference basis $\{|1\rangle,|2\rangle,|3\rangle\}$ if we have
all projectors ($P_j$) of rank one (i.e., $\{ P_1 = |1\rangle\langle 1|$,
$P_2 = |2\rangle\langle 2|$, $P_3 = |3\rangle\langle 3|\}$),  then the
corresponding PPIO would be a rank-one PPIO. However, if one of
the projectors is not rank-one (e.g., - $\{P_1 = |1\rangle\langle 1|
+ |2\rangle\langle 2|, P_2 = |3\rangle\langle 3|\}$) then the
corresponding PPIO would no more be a rank-one PPIO.

We now show that any local rank-one PPIO cannot increase correlated coherence,
which is similar to the action of local von Neumann measurement on the mutual information
(see Eq.~(14) in~\cite{Ollivier}).

{\it Theorem 1.}
Let $\check{\Pi}_a$ be any given local rank-one PPIO (on the subsystem $A$),
${\check{\Pi}^a}(\cdot):={\check{\Pi}_a}\otimes\mathbbm{1}_b(\cdot)$, then
\begin{eqnarray}
{I}_{co}[{\check{\Pi}^a}(\rho_{ab})]\leq {I}_{co}(\rho_{ab})
\label{coherencecorrelation2}
\end{eqnarray}
holds for any bipartite state $\rho_{ab}\in\mathcal{S}(H_a\otimes H_b)$.

{\it Proof.} Let $\check{\Pi}_a$ be a PPIO on part $A$.
Then, for any $\rho_{ab}$, we apply $\check{\Pi}^a$ on $\rho_{ab}$, the output state is
$\rho'_{ab}=(\check{\Pi}_a\otimes\mathbbm{1}_b)\rho_{ab}$. It follows that
\begin{eqnarray*}&&{I}_{co}(\rho_{ab})-{I}_{co}(\rho'_{ab})\\
&=&{C}_r(\rho_{ab})-{C}_r(\rho_a)-{C}_r(\rho_b)\\
&&-[{C}_r(\rho'_{ab})-{C}_r(\rho_a')-{C}_r(\rho_b')]\\
&=&{C}_r(\rho_{ab})-{C}_r(\rho_a)-[{C}_r(\rho'_{ab})-{C}_r(\rho_a')]\\
&=&S[\Delta(\rho_{ab})]-S(\rho_{ab})-S[\Delta(\rho_a)]+S(\rho_a)\\
&&-S[\Delta(\rho'_{ab})]+S(\rho'_{ab})+S[\Delta(\rho_a')]-S(\rho_a')\\
&=&S(\rho'_{ab})-S(\rho_{ab})+S(\rho_a)-S(\rho_a')+\Gamma(\rho_{ab},\check{\Pi}_a)\\
&=&I(\rho_{ab})-I(\rho'_{ab})+\Gamma(\rho_{ab},\check{\Pi}_a),
\end{eqnarray*}
where $\Gamma(\rho_{ab},\check{\Pi}_a)=S[\Delta(\rho_{ab})]-S[\Delta(\rho_a)]
-S[\Delta(\rho'_{ab})]+S[\Delta(\rho_a')]$.
If $\check{\Pi}_a$ is a rank-one PPIO with $U_i=U_j$
for any $i$ and $j$, it is clear since
$I(\rho_{ab})-I(\rho'_{ab})\geq0$ and $\Gamma(\rho_{ab},\check{\Pi}_a)=0$.
We now turn to the case of $U_i\neq U_j$ for some $i$ and $j$.
We assume with no loss of generality that
$\dim H_a=3$ and $\check{\Pi}_a(\cdot)=\sum_jU_jP_j(\cdot)P_jU_j^\dag$
with $U_1=|0\ra\la1|+|1\ra\la0|+|2\ra\la2|$, $U_2=U_3=I_3$.
It follows that, for any $\rho_{ab}=\sum_{i,j}|i\ra\la j|\otimes B_{ij}$,
where $B_{ij}\in\mathcal{B}(H_b)$,
\begin{eqnarray*}
\check{\Pi}^a(\rho_{ab})&=&|1\ra\la1|\otimes (B_{11}+B_{22})+|2\ra\la2|\otimes B_{33}\\
&=&p_1|1\ra\la 1|\otimes \rho'_{b,1}+p_2|2\ra\la 2|\otimes {\rho}'_{b,2},
\end{eqnarray*}
where $p_1{\rho}'_{b,1}=B_{11}+B_{22}$, $p_2{\rho}'_{b,2}=B_{33}$.
Let $B_{ii}=q_i{\varrho}'_{b,i}$, $i=1$, 2, 3,
$\Pi_a(\cdot)=\sum_jP_j(\cdot)P_j$ and $(\Pi_a\otimes \mathbbm{1}_b)\rho_{ab}=\varrho'_{ab}$.
Observing that
$S(\rho'_{ab})-S(\rho_a')=p_1S({\rho}'_{b,1})+p_2S({\rho}'_{b,2})$,
$S(\varrho'_{ab})-S(\varrho_a')=\sum_iq_iS({\varrho}'_{b,i})$,
and $\Gamma(\rho_{ab},\check{\Pi}_a)=q_1S[\Delta({\varrho}'_{b,1})]
+q_2S[\Delta({\varrho}'_{b,2})]-p_1S[\Delta({\rho}'_{b,1})]$, we thus have
\begin{eqnarray*}
&&I(\rho_{ab})-I(\rho'_{ab})+\Gamma(\rho_{ab},\check{\Pi}_a)\\
&=&I(\rho_{ab})-I(\varrho'_{ab})+I(\varrho'_{ab})-I(\rho'_{ab})+\Gamma(\rho_{ab},\check{\Pi}_a)\\
&=&I(\rho_{ab})-I(\varrho'_{ab})+p_1S({\rho}'_{b,1})-q_1S({\varrho}'_{b,1})-q_2S({\varrho}'_{b,2})\\
&&+q_1S[\Delta({\varrho}'_{b,1})]+q_2S[\Delta({\varrho}'_{b,2})]-p_1S[\Delta({\rho}'_{b,1})]\\
&=&I(\rho_{ab})-I(\varrho'_{ab})+q_1C_r(\varrho'_{b,1})+q_2C_r(\varrho'_{b,2})-p_1C_r(\rho'_{b,1})\\
&\geq&0
\end{eqnarray*}
since $I(\rho_{ab})-I(\varrho'_{ab})\geq0$ and
$q_1C_r(\varrho'_{b,1})+q_2C_r(\varrho'_{b,2})-p_1C_r(\rho'_{b,1})\geq0$, where
the later inequality holds since $C_r$ is convex \cite{Baumgratz2014prl}
(note that all the other nontrivial cases can be reduced to the convexity of $C_r$). We thus complete the proof.
\hfill$\blacksquare$

We remark here that, in general, for a PIO which is not a rank-one
PPIO, Eq.~(\ref{coherencecorrelation2})
is not true.
From the argument in the proof above, for any rank-one PPIO
$\check{\Pi}_a(\cdot)=\sum_jU_jP_j(\cdot)P_jU_j$,
let $\Pi^a(\cdot)=\sum_jP_j(\cdot)P_j$,
then we have
\begin{eqnarray}
I_{co}(\rho_{ab})-I_{co}[\check{\Pi}^a(\rho_{ab})]
\geq I(\rho_{ab})-I[(\Pi^a\otimes\mathbbm{1}_b)\rho_{ab}]~~~
\label{note}
\end{eqnarray}
and the equality holds whenever $\check{\Pi}^a(\rho_{ab})=\sum_ip_i|\pi(i)\ra\la\pi(i)|\otimes \rho_{b,i}$
for some permutation $\pi$ provided that $(\Pi^a\otimes\mathbbm{1}_b)\rho_{ab}=\sum_ip_i|i\ra\la i|\otimes \rho_{b,i}$.

We are now ready for giving our main concept, which is an analog
of quantum discord: mutual information $I$ is replaced by correlated
coherence $I_{co}$ and the von Neumann measurement is
substituted by rank-one PPIO.

{\it Definition 2.} Let $\rho_{ab}$ be a bipartite state in
$\mathcal{S}(H_a\otimes H_b)$. We define
\begin{eqnarray}
{D}^a_c(\rho_{ab}):
=\min\limits_{\check{\Pi}_a}\{{I}_{co}(\rho_{ab})-{I}_{co}[\check{\Pi}^a(\rho_{ab})]\},
\label{maincrrelation}
\end{eqnarray}
where the minimum is taken over all local rank-one PPIOs $\check{\Pi}_a$ and ${\check{\Pi}^a}(\cdot):={\check{\Pi}_a}\otimes\mathbbm{1}_b(\cdot)$.

One can readily verify that ${D}^a_c(\rho_{ab})=0$ if and only if
$\rho_{ab}=\sum_ip_i|i\rangle\langle i|\otimes \rho_i^b$.
By Theorem 3.1 in \cite{guo2012jpa}, ${D}^a_c(\rho)=0$ for
$\rho=\sum_{i,j}A_{ij}\otimes |i'\rangle\la j'|$ if
and only if $A_{ij}$s are mutually commuting normal
operators which are diagonal under the reference basis.
We denote the set of all states that with zero
${D}^a_c$ by $\mathcal{D}_{c-0}^a$. Then $\mathcal{D}_{c-0}^a$ is a convex set.
Let $\mathcal{D}_0^a$ be the set of all zero-discordant states (up to part $A$),
then $\mathcal{D}_{c-0}^a$ is a proper subset of $\mathcal{D}_0^a$.

Let $\{|y\rangle\}$ be the reference basis. An operation
$\mathcal{E}$ is called an incoherently unitary operation (IUO) if
\begin{eqnarray*}
\mathcal{E}(\rho)=U\rho U^\dag,\ U=\sum\limits_ye^{i\theta_y}|\pi(y)\rangle\langle y|.
\end{eqnarray*}
It is straightforward that ${D}^a_c$ is invariant under local IUO.
In addition, we can see that (i)
$D^a_c[\check{\Pi}^a(\rho_{ab})]=0$ for any local rank-one PPIO
$\check{\Pi}^a$ and any $\rho_{ab}$ and (ii) ${D}^a_c$ can be generated under LOCC.

We now discuss the relation between $D^a$ and $D^a_c$.
The following theorem is immediate from Eq.~(\ref{note}) and the proof of Theorem 1.
Since $D^a_c$ is defined to be the minimum over all $\check{\Pi}^a$ it also contains
the case $\check{\Pi}^a(\cdot)=\sum_j P_j(\cdot)P_j $ in which case
[along with other cases discussed after Eq.~(\ref{note})] the equality holds in Eq.~(\ref{note}).

{\it Theorem 2.} Let $\Omega$ be the set of all orthonormal bases of $H_a$. Then
for any $\rho_{ab}\in\mathcal{S}(H_a\otimes H_b)$,
\begin{eqnarray}
{D}^a(\rho_{ab})=\min_{\Omega}{D}^a_c(\rho_{ab}),
\label{relation of qd and qdc}
\end{eqnarray}
where the minimum is taken over all possible reference bases in $\Omega$.

Equation~(\ref{relation of qd and qdc}) displays the relation between
the quantum discord and the correlated coherence:
quantum discord is the minimal correlated bipartite coherence.
In addition, although the calculation of $D^a$ is NP-complete \cite{Huangyichen},
$D^a_c$ can be obtained directly since it is not dependent on the choice of the rank-one PPIO.
Furthermore, $D^a_c(\rho_{ab})$ not only displays the quantum discord contained in $\rho_{ab}$, but also
reflects the correlated coherence of $\rho_{ab}$. In other words,
$D^a_c$ reveals quantum correlation and coherence simultaneously.

Equation~(\ref{relation of qd and qdc})
is different from the Eq.~(16) in~\cite{Yaoyao}:
The discord in~\cite{Yaoyao} is not the
original quantum discord, it is the relative entropy of the discord,
and in addition, the basis-free measure of
coherence $\mathcal{C}^{\rm free}$ there is different from $D^a_c$
(since $\mathcal{C}^{\rm free}(\rho_{ab})=0$
iff $\rho_{ab}=\sum_{i,j}\delta_{ij}|i\ra|j'\ra\la i|\la j'|$,
where $\{|i\ra|j'\ra\}$ is the reference basis).
In addition, the correlated coherence was compared to the symmetric
quantum discord as well in~\cite{Tan}.
However, the result therein is far different
from Eq.~(\ref{relation of qd and qdc}): it is based on
$\mathcal{C}_{cc}$ while our relation is deduced from $D^a_c$.

\section{Relative entropy of the discordlike coherence}

We follow the axiomatic method for
a reasonable measure of quantum coherence ${C}(\rho)$
proposed in~\cite{Baumgratz2014prl}:
(C1) $C(\delta)\geq0$, $\forall$ $\delta\in \mathcal{I}$
[(C$1'$) $C(\delta)=0$ iff $\delta\in \mathcal{I}$].
(C2a) $C(\rho)$ is nonincreasing under incoherent operations,
i.e., $C[\mathcal{E}(\rho)]\leq C(\rho)$ for any IO $\mathcal{E}$.
(C2b) Monotonicity for average coherence under selective IO,
i.e., $C[\mathcal{E}(\rho)]\geq \sum_ip_iC(\rho_i)$,
with probabilities $p_i={\rm Tr}(K_i\rho K_i^\dag)$,
state $\rho_i=K_i\rho K_i^\dag/p_i$, and incoherent Kraus
operators $K_i$ obeying $K_i\mathcal{I}K_i^\dag\subseteq\mathcal{I}$.
(C3) Convexity, i.e., $\sum_jp_jC(\rho_j)\geq C(\sum_jp_j\rho_j)$ for any set
$\{p_j,\rho_j:\sum_jp_j=1\}$.

The relative entropy of the resource is an important quantity in
QRT \cite{Gilad2015prl}. We now discuss the relative
entropy of discordlike coherence. We define
\begin{eqnarray}
C^r(\rho_{ab}):=\inf\limits_{\sigma_{ab}\in\mathcal{D}_{c-0}^a}S(\rho_{ab}\|\sigma_{ab}).
\end{eqnarray}
Let $\sigma_{ab}=\sum_ip_i|i\ra\la i|\otimes\rho_i^b$ be
any state in $\mathcal{D}_{c-0}^a$.
It follows that $S(\rho_{ab}\|\sigma_{ab})=S[(\Delta\otimes\mathbbm{1}_b)\rho_{ab}]
-S(\rho_{ab})+S[(\Delta\otimes\mathbbm{1}_b)\rho_{ab}\|\sigma_{ab}]$, and thus we get
\begin{eqnarray}
C^r(\rho_{ab})=S[(\Delta\otimes\mathbbm{1}_b)\rho_{ab}]-S(\rho_{ab}).
\end{eqnarray}

It is clear that $C^r(\rho_{ab})$ fulfils (C1)(but it does not
satisfy (C1$'$), that is, $C^r(\rho_{ab})$ is not faithful).
It satisfies (C2a) and (C3) since the relative entropy
is contractive and jointly convex. It also fulfils (C2b)
by Theorem 5 in~\cite{Vedral} (also see the similar
argument as that of (C2b) for the original relative entropy
of coherence in \cite{Baumgratz2014prl}). That is, although
${D}^a_c$ is not a well-defined coherence measure since they
are not convex under the mixing of states, the relative entropy
measure $C^r$ is a measure of bipartite coherence.
The defect of this coherence measure is that it is not faithful,
which is similar to that of negativity \cite{Vidal1} as an entanglement measure.
So we can also consider the symmetric discordlike measure and
the symmetric relative entropy of the discordlike quantity,
denoted by $\tilde{{D}}_c$ and $\tilde{C}^r$, respectively. That is,
\begin{eqnarray}
\tilde{{D}}_c(\rho_{ab}):
=\min\limits_{\check{\Pi}_a\otimes\check{\Pi}_b}\{{I}_{co}(\rho_{ab})
-{I}_{co}[(\check{\Pi}_a\otimes\check{\Pi}_b)\rho_{ab}]\},
\label{maincrrelation3}
\end{eqnarray}
where the minimum is taken over all local PPIOs
$\check{\Pi}_a\otimes\check{\Pi}_b$. For this symmetric
measure, $\tilde{{D}}_c(\rho_{ab})=0$ if and only if
$\rho_{ab}$ is diagonal with respect to the reference basis $\{|i\rangle|j\ra\}$.
Let $\tilde{\mathcal{D}}_{c-0}^a$ be the set of all states with
zero symmetric discordlike measure. The corresponding relative entropy measure is
\begin{eqnarray}
\tilde{C}^r(\rho_{ab}):
=\inf\limits_{\sigma_{ab}\in\tilde{\mathcal{D}}_{c-0}^a}S(\rho_{ab}\|\sigma_{ab}).
\end{eqnarray}
That is, the symmetric measure of relative entropy coincides
with the original relative entropy coherence measure.

\section{QRT of the discordlike coherence}

In what follows, we discuss whether $D^a_c$ obeys the requirements of QRT.
Let $\mathcal{F}$ be the set of all free states (in all possible finite dimensions),
and $\mathcal{F}_m=\mathcal{F}\cap\mathcal{S}(H_m)$,
where $H_m=H_{m_1}\otimes H_{m_2}\otimes\cdots H_{m_s}$,
$\dim H_{m_i}=m_i$, $i=1$, $\dots$, $s$. Brand\~{a}o and Gour proposed
the following postulates \cite{Gilad2015prl} for any QRT:
(i) $\mathcal{F}$ is closed under tensor products; (ii) $\mathcal{F}$ is
closed under the partial trace of spatially separated subsystems;
(iii) $\mathcal{F}$ is closed under permutations of spatially separated subsystems;
(iv) each $\mathcal{F}_m$ is a closed set; (v) each $\mathcal{F}_m$ is a convex set;
(vi) the set of free operations cannot generate a resource;
they cannot convert free states into resource states.

When we process a quantum task, the interacting environment always
consumes resources. To account for this very recently Chitambar and Gour
investigated the physically consistent QRT \cite{Gilad2016prl}.
A QRT defined on some quantum system $S$ is physically consistent if any free operation
$\mathcal{E}$ on $S$ can be obtained by an auxiliary state $\rho_E$, a joint unitary
$U_{S,E}$, and a projective measurement $\{P_i\}$ that are all free
in an extended system $S+E$. We call the free operation in the
physically consistent QRT physically free operation hereafter.
The following is the main result of this section.

{\it Theorem 3.} A quantum operation
$\mathcal{E}:\mathcal{B}(H_a\otimes H_b)\rightarrow\mathcal{B}(H_a\otimes H_b)$
is a physically free operation of the discordlike
correlation of coherence if and only if it can be expressed as a
convex combination of maps each having Kraus operators $\{K_j\}_{j=1}^r$ of the form
\begin{eqnarray}
K_j=U_a\otimes B_{j},
\label{physically free operation of qd}
\end{eqnarray}
where $U_a$ is an IUO on $H_a$ and $\{B_{j}\}_{j=1}^r$
is any Kraus operators that satisfy $\sum_jB_{j}^\dag B_{j}=I_b$.

{\it Proof.} We assume that the environment is denoted by part E.
Following the scenario in~\cite{Gilad2016prl},
any physically free operation of the discordlike correlation of coherence
on this composite system can be decomposed into three steps:
(i) a joint unitary $U_{ab,E}$ is applied on the input
state $\rho_{ab}$ and some fixed state $\rho_E$, i.e.,
the state becomes $U_{ab,E}\rho_{ab}\otimes \rho_EU_{ab,E}^\dag$,
where $U_{ab,E}=U_a\otimes U_{bE}$, $U_a$ is a IUO
(note that the joint unitary operation is free,
so it admits this form)
 and $U_{bE}$
is an arbitrary given unitary operator on $H_b\otimes H_E$,
(ii) a von Neumann measurement $\{P_j\}$ act on the environment
encoding the measurement outcome as a classical index, i.e.,
the state after this process is
\begin{eqnarray*}
&&\sum_j(I_{ab}\otimes P_j)(U_{ab,E}\rho_{ab}\otimes \rho_EU_{ab,E}^\dag)(I_{ab}\otimes P_j)\\
&=&\sum_{j=1}^t\rho_{ab,j}\otimes|j\rangle_E\langle j|,
\end{eqnarray*}
where
\begin{eqnarray*}
\rho_{ab,j}={\rm Tr}_E[(I_{ab}\otimes P_j)
(U_{ab,E}\rho_{ab}\otimes \rho_EU_{ab,E}^\dag)],
\end{eqnarray*}
and (iii) a classical processing channel is applied to
the measurement outcomes. That is, the final state is
$\sum_{k=1}^{t'}\rho_{ab,k}'\otimes|k\rangle_E\langle k|$,
where $\rho_{ab,k}'=\sum_{j=1}^{t}p_{k|j}\rho_{ab,j}$
for some classical channel $p_{k|j}$.

From the discussion above, let
$\rho_{ab}=\sum_{ij}|i\rangle\langle j|\otimes B_{ij}$
with respect to the reference basis $\{|i\ra\}$ of $H_a$,
then we can conclude that
\begin{eqnarray*}
\rho_{ab,j}&=&{\rm Tr}_E[(I_{ab}\otimes P_j)(U_a\otimes U_{bE}\rho_{ab}
\otimes \rho_EU_a^\dag\otimes U_{bE}^\dag)]\\
&=&{\rm Tr}_E\left[\sum_{kl}U_a|k\ra\la l|U_a^\dag\otimes I_{b}
\otimes P_jU_{bE}(B_{kl}\otimes \rho_E)U_{bE}^\dag\right]\\
&=&\sum_{kl}U_a|k\ra\la l|U_a^\dag\otimes {\rm Tr}_E[I_{b}
\otimes P_jU_{bE}(B_{kl}\otimes \rho_E)U_{bE}^\dag].
\end{eqnarray*}
Then the final state becomes
\begin{eqnarray*}
&&\rho'_{ab}=\sum_j\rho_{ab,j}\\
&=&\sum_j\sum_{kl}U_a|k\ra\la l|U_a^\dag\otimes {\rm Tr}_E[I_{b}
\otimes P_jU_{bE}(B_{kl}\otimes \rho_E)U_{bE}^\dag]\\
&=&\sum_{kl}U_a|k\ra\la l|U_a^\dag\otimes {\rm Tr}_E[U_{bE}(B_{kl}\otimes \rho_E)U_{bE}^\dag],
\end{eqnarray*}
which completes the proof since any quantum operation on the system $B+E$
admits the form of ${\rm Tr}_E[U_{bE}(\cdot)U_{bE}^\dag]$.
\hfill$\blacksquare$

By Theorem 3, one can readily obtain that a quantum operation
$\mathcal{E}$ is a physically free operation for quantum discord
if and only if it has the form as Eq.~(\ref{physically free operation of qd}) with
$U_{a}$ is any unitary operation on $H_a$ since the difference between $D^a_c$ and $D^a$
is in nature the fixed reference basis for $D^a_c$ comparing with the free bases for $D^a$.

When we regard $\mathcal{D}_{c-0}^a$ as the free states, and consider
the operations in Theorem 3 as physically free operations, we show below that
it is a new physically consistent resource under QRT.

Since ${D}^a_c$ is a measure of bipartite systems,
we assume that $H_{m_i}=H_{a_i}\otimes H_{b_i}$, $i=1$, $\dots$, $s$,
and $\rho\in\mathcal{S}(H_m)$ is free if any reduced state in $H_{m_s}$ is free.
We check the postulates (i) through (v) for free states item by item.
(i) With no loss of generality, we consider the case of $s=1$.
If $\rho_{ab}\in\mathcal{S}(H_a\otimes H_b)$ with ${D}^a_c(\rho_{ab})=0$,
then $\rho_{ab}=\sum_{i}p_i|i\ra\la i|\otimes\rho_i^b$ with respect to the reference
basis $\{|i\ra\}$ of $H_a$. It follows that
$\rho_{ab}\otimes\rho_{a'b'}\in\mathcal{S}(H_a\otimes H_b\otimes H_{a'}\otimes H_{b'})$
can be expressed as $\rho_{aba'b'}=(\sum_{i}p_i|i\ra\la i|\otimes\rho_i^b)
\otimes(\sum_{i}p_i'|i'\ra\la i'|\otimes\rho_i^{b'})$ and thus it
is a free state in $\mathcal{S}(H_a\otimes H_{a'}\otimes H_b\otimes H_{b'})$.
(ii) and (iii) are clear. It is easy to check (iv) according to Theorem 3.1 in \cite{guo2012jpa}.
Postulate (v) is clear and the properties for the free operations are guaranteed by Theorem 3.
We thus conclude that the discordlike correlation of bipartite coherence
can also be regarded as a quantum resource
(the comparison between ${D}^a_c$ and ${D}^a$ is listed in Table~\ref{tab:table1}).

It is worth mentioning that for both entanglement and coherence,
the free states are closely related to the free operations.
A state is entangled if and only if it can not be prepared by
LOCC, and a state is incoherent if and only if
it can not be created via incoherent operations \cite{Baumgratz2014prl}.
However, quantum discord is defined via
local von Neumann measurements, which is only a proper set of
its free operations (one can see that any quantum operation
$\mathcal{E}:\mathcal{B}(H_a\otimes H_b)\rightarrow\mathcal{B}(H_a\otimes H_b)$
that has the Kraus operators of the form
$K_j=A_j\otimes B_j$ is a free operation of quantum discord,
where $\sum_jA_j(\cdot)A_j^\dag$ is a commutativity preserving
operation \cite{explain} and $\sum_jB_j(\cdot)B_j^\dag$
is any quantum operation on part $B$). More remarkably, the free
states of quantum discord is not convex. All these facts indicates
that quantum discord as a quantum resource is not a ``good'' candidate.

\section{Conclusion}

Quantum discord has two indispensable shortcomings:
it is difficult to compute and it
does not obey the structure of QRT.
But the closely related correlation,
the discordlike correlation of coherence we established here,
overcomes these defects completely.
We show that the minimal discordlike correlation of coherence over
all possible reference bases turns out to be exactly the quantum discord of
the bipartite state (thus it vastly improves both Theorem 2 in
~\cite{Yaoyao} and the results in~\cite{Tan}).
This indicates the inherent nature of quantum discord as
a bipartite coherence. Moreover, the discordlike correlation of coherence
can be calculated in a straightforward manner and
obeys all requirements of QRT. Interestingly, we also prove
that the relative entropy measure induced from this measure can
be presented as a coherence measure of the bipartite system. It is far
different from all the previous ways of quantifying
coherence in nature since ${D}^a_c$ is induced via measurements.
We believe that the discordlike correlation of coherence would
be more useful than the original quantum discord
as a quantum resource in quantum information technology.

\begin{acknowledgements}
This work was completed while Guo was visiting
the Institute of Quantum Science and
Technology of the University of Calgary
under the
support of the China Scholarship Council
(Grant No. 201608140008).
Guo thanks Professor
C. Simon, Professor A. Lvovsky
and Professor G. Gour for their hospitality.
The authors thank Professor S. Wu for helpful discussions.
We also would like to thank an anonymous referee for
useful and important suggestions.
Guo is supported by the National Natural
Science Foundation of China under Grant No. 11301312 and the Natural Science Foundation of Shanxi
under Grants No. 201701D121001.
\end{acknowledgements}

\end{document}